\documentclass[twocolumn]{revtex4}
\usepackage{graphicx}

\usepackage[]{epsfig}
\usepackage{verbatim} 
\usepackage{amssymb}
\newcommand{\beq}{\begin{equation}}
\newcommand{\eeq}{\end{equation}}
\newcommand{\beqd}{\begin{displaymath}}
\newcommand{\eeqd}{\end{displaymath}}
\newcommand{\beqa}{\begin{eqnarray}}
\newcommand{\eeqa}{\end{eqnarray}}

\newcommand{\e}{\epsilon}

\newcommand{\s}{\sigma}
\newcommand{\A}{{\mathcal A}}

\begin{document}

\title{On Cavity Approximations for Graphical Models}

%[pre,aps,showpacs]

\author{T. Rizzo\dag, B. Wemmenhove\ddag \ and H.J. Kappen\ddag}
\affiliation{
\dag ``E Fermi'' Center, Via Panisperna 89A, Compendio del Viminale 00184, 
Rome, Italy
\\
\ddag Dept. of Biophysics, Foundation for Neural Networks (SNN) \\
Geert Grooteplein 21, 6525 EZ Nijmegen, Netherlands}

\begin{abstract}
We reformulate the Cavity Approximation (CA), a class of algorithms recently introduced for improving the Bethe approximation estimates of marginals in graphical models. In our new formulation, which allows for the treatment of 
multivalued variables, a further generalization to factor graphs with 
arbitrary order of interaction factors is explicitly carried out, and a 
message passing algorithm that implements the first order correction to
the Bethe approximation is described.
Furthermore we investigate an implementation of the CA for pairwise 
interactions.
In all cases considered we could confirm that CA[k] with increasing $k$
 provides
a sequence of approximations of markedly increasing precision. Furthermore in
some cases we could also confirm the general expectation that the 
approximation of order $k$,
whose computational complexity is $O(N^{k+1})$ has an error that scales as
$1/N^{k+1}$ with the size of the system.
We discuss the relation between this approach and some recent developments in the field.
\end{abstract}

\maketitle
\section{Introduction}
The Bethe approximation (BA) is one of the major ingredients leading to the important advances in combinatorial optimisation made by the statistical physics community in recent years.
The starting point of this line of research can be traced back to {\it a)} the inclusion the Replica-Symmetry-Breaking (RSB) scheme in the context of the Bethe approximation  \cite{MP1,MP2} and {\it b)} the application of the method to single instances \cite{MPZ}.
On the other hand the Bethe approximation has become a key issue in the context of information theory after 
it was recognized that the well known Belief-Propagation (BP) algorithm is tightly related to it \cite{YWF}. 
This algorithm was introduced in the context of Bayesian networks and has gained interest  after the 
discovery that the fast decoding of Turbo codes and Gallagher codes is indeed an instance of BP \cite{MMC}. 
Currenlty the problem of computing the corrections to the BA is attracting incresing attention (see \cite{MR,CC,PS} for recent literature,) not only for the applications mentioned above but also because the BA is the only way of obtaining a mean-field like solution to many unsolved physical problems, notably Anderson localization.

In this work we re-investigate the Cavity Approximation (CA), a tool recently introduced in \cite{MR} to study graphical models.
The CA is a sequence of approximations defined iteratively such that the BA corresponds to the zero-th order. Its main features are:
i) it can be implemented on a given sample (much as the Bethe approximation and at variance with the Replica method), therefore to each approximation corresponds a BP-like algorithm; ii)
the expansion at order $k$ (CA[$k$]) is correct on graphs with $k$ loops, much as the Bethe approximation is correct on trees;  iii) the computational complexity of the corresponding algorithm grows as $N^{k+1}$;
 iv) when averaged over the samples the CA reproduces the results of the Replica method, indeed it corresponds to computing the $1/N^{k}$ corrections within the cavity method. 
In \cite{MR} it was argued that the CA
 is the natural approximation scheme on locally tree-like structures, in the sense that  CA[$k$] yields the $O(1/N^k)$ corrections for models defined on random graphs. 
In this paper we confirm this expectation by implementing algorithimically the CA, in particular we apply  CA[0] ({\it i.e.} BA), CA[1] and CA[2] to instances of graphical models defined on random graphs. We conclude that the CA is an efficient tool to improve (with polynomial complexity) the BA on this class of models that includes notably the error-correcting codes mentioned above. We also formulate the theory in a representation that allows for straightforward generalization to factor graphs with arbitrary order of the interaction factors. Message passing equations for the implementation of such a generalization are given explicitly. We discuss the relationship between this approach and other approaches to go beyond the BA. 

\section{The Cavity Approximation: basic ideas}
In \cite{MR} the approach was illustrated in the case of binary
variables with pairwise interactions. In the following, for the sake of completeness, we present the case of
multivalued variables with generic pairwise interactions $H_{ij}(x_i,x_j)$. The same ideas and methods can be applied to models with multiple interactions (factor graphs).

\begin{figure}[htb]
\epsfig{file=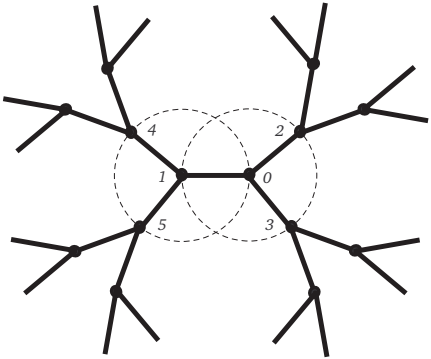,width=6cm} 
\caption{The marginals of nodes $0$ and $1$ in the absence of link
$(01)$ can be expressed in terms of the joint probabilities of nodes
$0,4,5$ in the absence of node $1$ or of the joint probabilities of nodes 
$1,2,3$ in the absence of node $0$. The equality of the results yields the
cavity equations}
\label{figure1}
\end{figure}

The basic assumption of the BA is that, once a node (say $\s_0$ in fig.
\ref{figure1}) is removed from the system, the nodes that were connected to it
($\s_1$,$\s_2$ and $\s_3$) become uncorrelated. This is true on a tree but
it is not true in general if loops not shown in fig.
\ref{figure1} are present. From this assumption one can obtain estimates of local averages
of the variables. We
consider two questions:
\begin{enumerate} 
\item How can we estimate the correlation between node $\sigma_2$ and
$\sigma_3$ when node $\sigma_0$ is removed from the system?
\item How can we use these correlations to improve the estimates of the local
averages?   
\end{enumerate}
In order to answer these questions local cavity distributions are introduced and
equations are derived for them. 
The equations will not be sufficient to compute all the cavity distributions and
 they will be partially estimated through a Bethe-like approximation.
For each node $i$ we define $\partial i$ the neighbors of $i$ and $x_{\partial
i}\equiv \{ x_j : j \in \partial i\}$. 
For each node $i$ we consider its cavity distribution, defined as the
distribution $P^{(i)}(x_{\partial i})$ of its neighbors in the graph, obtained
by removing the variable $i$ from the original graph. 
Note that the knowledge of $P^{(i)}(x_{\partial i})$ (the Markov blanket of $i$)
is sufficient to determine $P(x_i,x_{\partial i})$ through the formula:
\beq
P(x_i,x_{\partial i})=c  P^{(i)}(x_{\partial i})\prod_{j \in \partial
i}\psi_{ij}(x_i,x_j)
\eeq 
where $\psi_{ij}(x_i,x_j)\equiv \exp(-\beta H_{ij}(x_i,x_j))$ and $c$ is a
normalization constant.
Now we consider the effect of adding to the system without node $x_0$ 
all the interactions but $\psi_{10}$.  We can express the marginal of
site $x_0$ in this system  in terms of $P^{(0)}(x_{\partial 0})$:
\begin{eqnarray}
P^{(01)}(x_0)  =  c \left(\sum_{ \{  x_{\partial 0/1} \} } \prod_{j \in
\partial
0/1}P^{(0)}(x_j)\psi_{0j}(x_0,x_j) \right. \nonumber \\
\left. +\sum_{ \{  x_{\partial 0/1} \}
}\epsilon^{(0)}( x_{\partial 0/1}) \prod_{j \in
\partial
0/1}\psi_{0j}(x_0,x_j)  \right)
\label{ce1}
\end{eqnarray}
where $c$ is a normalization and we have introduced the connected cavity
correlation of the set $x_{\partial 0/1}$, $\epsilon^{(0)}( x_{\partial
0/1})\equiv P^{(0)}( x_{\partial 0/1})-\prod_{j \in
\partial
0/1}P^{(0)}(x_j)$.
The same object may be calculated starting from the system without the variable
node $x_1$ and inserting all interactions but $\psi_{10}(x_1,x_0)$:
\begin{eqnarray}
P^{(01)}(x_0)  &=&      P^{(1)}(x_0)+  \qquad \qquad \nonumber \\
&& \hspace*{-20mm}
 \frac{\sum_{\{ x_1,x_{\partial 1/0} \}}
\e^{(1)}(x_0,x_{\partial 1/0})\prod_{i \in \partial 1/0} \psi_{1j}(x_1,x_j) }
{\sum_{\{ x_1,x_{\partial 1/0} \}}
P^{(1)}(x_{\partial 1/0})\prod_{i \in \partial 1/0} \psi_{1j}(x_1,x_j)}
\label{ce2}
\end{eqnarray}
where we have introduced another cavity connected correlation $\epsilon^{(1)}(
x_0,x_{\partial 1/0})\equiv P^{(1)}(x_0,
x_{\partial 1/0})-P^{(1)}(x_0)P^{(1)}(x_{\partial 1/0})$ and the suffix means
that quantities are computed in the system without node $x_1$.
Equating the r.h.s.'s of eq. (\ref{ce1}) and (\ref{ce2}), we obtain an  
equation
that connects the cavity distributions of neighboring nodes:
\beqa
P^{(1)}(x_0) & = & c \left(\sum_{ \{  x_{\partial 0/1} \} } \prod_{j \in
\partial
0/1}P^{(0)}(x_j)\psi_{0j}(x_0,x_j) \right. \nonumber \\
&& \left. + \sum_{ \{  x_{\partial 0/1} \}
}\epsilon^{(0)}( x_{\partial 0/1}) \prod_{j \in
\partial
0/1}\psi_{0j}(x_0,x_j)  \right)
\nonumber \\
 &&\hspace*{-15mm} - \frac{\sum_{\{ x_1,x_{\partial 1/0} \}}
\e^{(1)}(x_0,x_{\partial 1/0})\prod_{i \in \partial 1/0} \psi_{1j}(x_1,x_j) }
{\sum_{\{ x_1,x_{\partial 1/0} \}}
P^{(1)}(x_{\partial 1/0})\prod_{i \in \partial 1/0} \psi_{1j}(x_1,x_j)}
\label{exactce}
\eeqa
We note that this equation is {\it exact} and is valid also if some of the  nodes
connected to $x_0$ coincide with those connected to $x_1$.
We have $2 L$ such equations, two for each link (the
other equation for link $(01)$ is obtained   exchanging indices in eq.
(\ref{exactce}) according to $\{0  \leftrightarrow  1,2\leftrightarrow
4,3\leftrightarrow 5\}$).
Unfortunately these equations are not sufficient to determine the full set
of cavity distributions, which is easily seen noticing that if we knew all the
connected cavity correlations $\e^{(i)}(x_j,x_{\partial i/j})$ and
$\epsilon^{(j)}( x_{\partial j/i})$ for each link $(i,j)$ then the $2 L$ cavity
equations should be in principle sufficient to determine the remaining $2 L$
unknown cavity distributions $P^{(j)}(x_i)$.
The Bethe approximation assumes that the variable nodes on the cavity of node
$i$ are uncorrelated in the absence of node $i$. As a consequence the corresponding probability distributions are factorized and 
the connected correlations are zero ($\e^{(i)}(x_j,x_{\partial i/j})=0$, 
$\epsilon^{(j)}( x_{\partial j/i})=0$) for each link $(i,j)$, therefore eq.
(\ref{exactce}) reduces to the standard Belief-Propagation equation.

\section{Estimating the cavity distribution}

In general, if we have an estimate of $P^{(j)}(x_{\partial j})$ for any node $j$ we
can compute the various connected correlations in eq. (\ref{exactce})
and solve the cavity equations obtaining a new estimate of $P^{(i)}(x_j)$.
In the following we argue that to estimate the joint probability distribtution $P^{(j)}(x_{\partial j})$ it is sufficient to
have an algorithm ({\it e.g.} BP) that estimates single site marginals $P(x_i)$.
Indeed suppose that we have such an algorithm, then in order to
get an estimate of $P^{(j)}(x_{\partial j})$ we remove node $j$ from the graph
and evaluate $P^{(j)}(x_{j_1})$ through the given algorithm, where $\{
x_{j_1},\dots,x_{j_k} \}\equiv \partial j$. Then we fix the value of $x_{j_1}$
and compute $P^{(j)}(x_{j_2}|x_{j_1})$ through the same algorithm, and so on.
In the end the distribution can be reconstructed from the formula:
\beq
P^{(j)}(x_{\partial j})=
P^{(j)}(x_{j_1})\prod_{i=2}^{k}P^{(j)}(x_{j_i}|x_{j_1},\dots,x_{j_{i-1}})
\label{eqitP}
\eeq
where $k$ is the number of nodes on the cavity of $j$. 
In other words, in order to determine  $P^{(j)}(x_{\partial j})$ we have to run the
approximate algorithm removing site $j$ and fixing sequentially the values of
$x_{\partial j}$. Note that this procedure to estimate the cavity connected correlations by sequentially fixing the values of the cavity spins is easier to implement than the use of the Fluctuation-Dissipation-Theorem originally proposed in \cite{MR} since the latter requires taking derivatives of the eqs. (\ref{exactce}).

Any algorithm may be used to obtain a first estimate of $P^{(j)}(x_{\partial j})$, in particular we can use the BA and obtain a new cavity approximation of order 1 (CA[1]).
The procedure can be iterated yielding CA[$k$] (with CA[0]$\equiv$BA):
{\tt \begin{enumerate}
\item write the exact cavity equations for the system
\item for each variable node $i$:
\begin{enumerate}
\item remove $x_i$
\item express $P^{(i)}(x_{\partial i})$ 
in terms of conditional probabilities through eq.(\ref{eqitP})
\item use CA[$k-1$] to compute the conditional probabilities, compute\\
$P^{(i)}(x_{\partial i})$ and then $\epsilon^{(i)}$ 
\end{enumerate}
\item substitute the estimates of $\epsilon^{(i)}$ into the exact equations and
recompute \\ the $2L$ cavity distributions $P_j^{(i)}(x_j)$.
\end{enumerate}}
In practice the procedure can be implemented through a message-passing algorithm
whose computational complexity grows with order $k$ as $N^{k+1}$. 

\section{Perturbative approach for practical implementations}
\label{perturbative}
We note that  the use of eq. (\ref{eqitP}) requires the application of the algorithm 
CA[$k-1$] a number of times exponential in the size of the cavities, therefore it
may be convenient to use an approximate expression of eq. (\ref{eqitP}). In the following we discuss 
one such approximation.
For a given set of nodes ${\mathcal A}$ we define the connected correlation functions as usual, in particular we have:
$c(x)=P(x)$, $c(x,y)=P(x,y)-P(x)P(y)$ and so on. The probability distribution
of a set of nodes $\A$ can be written as:
\beq
P(x_{\A})=\sum_{[\A_1,\dots,\A_n]}c(x_{\A_1})\dots c(x_{\A_n})
\label{eqitP2}
\eeq
where $[\A_1,\dots,\A_n]$ runs over the partitions of $\A$.
Under some conditions one can assume that $P(x)$ is $O(1)$ while $c(x,y)$ is small,
say $O(\epsilon)$ (where $\epsilon$ is some small parameter),
$c(x,y,z)=O(\epsilon^2)$ and so on.
For instance in the representation  
$P(x_\A)\propto \exp[\sum_i
a_i(x_i)+\sum_{i<j}a_{ij}(x_i,x_j)+\sum_{i<j<k}a_{ijk}(x_i,x_j,x_k)+\dots]$ this
approximation is valid if the interaction terms between $k$ variables are
proportional to $\epsilon^{k-1}$.
As before the connected correlation functions can be expressed 
through conditional probabilites, {\it i.e.} $c(x,y)=(P(x|y)-P(x))P(y)$
and can be determined through any  algorithm that yields the local
distributions $P(x)$.
These observations can be used to reduce the number of quantities to be estimated at each 
cavity, in particular, steps {\tt 2.b} and {\tt 2.c} can be modified in the following way:
{\tt \begin{itemize}
\item (b) express $P^{(i)}(x_{\partial i})$ through eq. (\ref{eqitP2})
assuming that all connected\\ correlation functions of more than $k+1$ nodes
vanish.
\item (c) use CA[$k-1$] on the corresponding system to
 determine the connected \\ correlations through conditional probabilities.
\end{itemize}}
In the following we call CA[$k$] the approximation scheme that includes the
previous assumption.
It was shown in \cite{MR} that CA[$k$] is exact on graphs with $k$ loops, much
as the Bethe approximation is exact on trees. 
It can be argued that this approximation scheme
yields the perturbative expansion in powers of $1/N$ on models defined on random
graphs of size $N$, roughly speaking it means that the CA[$k$] yields the local
marginals with an error $O(1/N^{k+1})$.
Indeed in the large $N$ limit random graphs are locally tree-like, the loops typically being large.
On a locally tree-like portion of a random graph the $2$-point cavity
connected correlations are determined by large loops and therefore are small;
the
$3$-point cavity correlations depend on the correlations between these large
loops and are even smaller, in general we expect
that the cavity correlations of $k$ nodes yield an effect $O(1/N^{k-1})$. 
Therefore
in such a region we expect that CA[$k$] is really a perturbative expansion.
On the other hand small loops  ($l \ll \ln N$) are definitely present in
random graphs, see \cite{MM} and Ref. therein.
The typical graph contains a finite number of small loops and in general graphs
with a finite number of small structures of $k$ nearby loops have probability
$O(1/N^{k-1})$.
Using the exactness of CA[$k$] on graphs with $k$ loops \cite{MR} mentioned
above 
it can be argued that the presence of these small loops does not destroy the
perturbative nature of the expansion. 

\section{Generalization to arbitrary factor graphs.}
The above strategy, which up to now has been restricted to 
two-variable interaction models,  
may be generalized surprisingly easily for factor graphs with arbitrary
number of variables in each factor.
We will write down exact equations, as before, for a definition of epsilon
functions that corresponds to an expansion
around totally factorizing cavity distributions, and later we will 
neglect higher order terms. The resulting equations explicitly yield a
message passing algorithm that takes into account the first order 
correction (CA[$1$]) to belief propagation. 
In our notation, Roman indices ($i,j,k,\ldots$)
will denote variables and Greek indices ($\alpha, \beta, \gamma, \ldots$) 
denote factors. The factor indices are understood to
simultaneously represent the subset of 
Roman indices corresponding to the variables in the factor.
\subsection{Exact equations}
The exact equations for the marginal of variable $x_i$ in the absence
of factor $\alpha$ reads
\begin{eqnarray}
P^{(\alpha)}(x_i)& = &c  \left( \sum_{x_{\partial_i \setminus \alpha}}
\prod_{\beta \in \partial_i \setminus \alpha}
\left[ \psi_\beta
(x_\beta) \prod_{j\in \beta \setminus i} P^{(i)}(x_j)\right] 
\right. \nonumber \\
&&\left.
\hspace*{-10mm}
+ \sum_{x_{\partial_i \setminus \alpha}} \epsilon^{(i)}(x_{\partial_i
\setminus \alpha})\prod_{\beta \in \partial_i \setminus \alpha}
\psi_\beta(x_\beta) \right) \\
P^{(\alpha)}(x_i)&=& P^{(j)}(x_i) \nonumber \\
&& \hspace*{-20mm}+ \frac{\sum_{x_j,x_{\partial_j \setminus \alpha}}
\epsilon^{(j)}(x_i,x_{\partial_j \setminus \alpha})
\prod_{\beta \in 
\partial_j \setminus \alpha} \psi_\beta(x_\beta)}
{
\sum_{x_j,x_{\partial_j \setminus \alpha}}
P^{(j)}(x_{\partial_j \setminus \alpha})\prod_{\beta \in 
\partial_j \setminus \alpha} \psi_\beta(x_\beta)}
\end{eqnarray}
where the expansion parameters are given by
\begin{eqnarray}
\epsilon^{(i)}(x_{\partial_i \setminus \alpha})& =& P^{(i)}(x_{\partial_i
\setminus \alpha})
 - \! \! \prod_{\beta \in \partial_i \setminus \alpha}
[\prod_{l\in \beta \setminus i} P^{(i)}(x_l)] \ \  \\
\epsilon^{(j)}(x_i,x_{\partial_j \setminus \alpha})&=&
P^{(j)}(x_i,x_{\partial_j \setminus \alpha})
\nonumber \\
&&-P^{(j)}(x_i)P^{(j)}(x_{\partial_j
\setminus \alpha})
\end{eqnarray}
\subsection{Truncated expansion}
In the following, we will assume that 
$\forall i \forall \alpha_1, \alpha_2 \in \partial_i$ we have that 
$\alpha_1 \cap \alpha_2 = \{i\}$.
Up to first order in two-point connected
correlations, we may write
\begin{eqnarray}
\epsilon^{(j)}(x_i,x_{\partial_j \setminus \alpha}) 
& \approx &\sum_{\beta \in \partial_j \setminus \alpha} \sum_{k\in 
\beta \setminus i}
c^{(j)}(x_i,x_k) \qquad\nonumber \\
&&  \hspace*{-20mm}
\times \prod_{l\in \beta \setminus i,k}
[P^{(j)}(x_l)] \prod_{\gamma \in \partial_j \setminus
(\alpha, \beta)} [\prod_{m\in \gamma \setminus j}P^{(j)}(x_m)]\\
\epsilon^{(i)}(x_{\partial_i \setminus \alpha}) & \approx & 
\sum_{\beta \in \partial_i \setminus \alpha} \sum_{k<l \in \beta \setminus i}
c^{(i)}(x_k,x_l)
\nonumber \\ && \hspace*{-20mm}\times \prod_{m\in \beta \setminus (k,l,i)}
[P^{(i)}(x_m)]
\prod_{\gamma \in \partial_i \setminus (\alpha, \beta)} [\prod_{n \in
\gamma \setminus i} P^{(i)}(x_n)]\nonumber \\
&& \hspace*{-20mm}
+ \sum_{\beta < \gamma \in \partial_i \setminus \alpha}
\sum_{k \in \beta} \sum_{l \in \gamma} c^{(i)}(x_k,x_l)
\nonumber \\
&&
\times \prod_{m\in \beta \setminus (i,k)}[P^{(i)}
(x_m)]\prod_{n\in \gamma \setminus (i,l)}[P^{(i)}(x_n)] 
\nonumber \\ && \hspace*{-20mm} \times
\prod_{\eta \in \partial_i \setminus (\alpha,\beta, \gamma)}
\prod_{r\in \eta \setminus i}[P^{(i)}(x_r)]
\end{eqnarray}
Let us introduce some notation:
\begin{eqnarray}
\mu_{\alpha \to i}(x_i) & \equiv & \sum_{x_{\alpha \setminus i}}
\prod_{j \in \alpha \setminus i}[ P^{(i)}(x_j)] \psi_\alpha(x_\alpha) \\
\lambda_{\alpha \to i}(x_i) & \equiv & \sum_{x_{\alpha \setminus i}}
\sum_{j < k \in \alpha \setminus i}
c^{(i)}(x_{j},x_{k}) \nonumber \\
&& \times \prod_{l \in \alpha \setminus (i,j,k)}[P^{(i)}(x_l)]
\psi_\alpha(x_\alpha) \\
\rho_{(\alpha,\beta) \to i}(x_i) & \equiv & \sum_{x_{\alpha \setminus i}}
\sum_{x_{\beta \setminus i}}\sum_{j \in \alpha \setminus i}
\sum_{ k \in \beta \setminus i}c^{(i)}(x_j,x_k) \nonumber \\
&& \hspace*{-20mm} \times \psi_\alpha(x_\alpha) \psi_\beta(x_\beta)
\prod_{l\in \alpha \setminus
(i,j)}[P^{(i)}(x_l) ]\prod_{m\in \beta \setminus (i,k)}[P^{(i)}(x_m)]\nonumber
 \\
\ \\
\nu_{\alpha \to i}(x_i,x_j) & \equiv & \sum_{x_{\alpha \setminus i}}
\sum_{k \in \alpha \setminus i} c^{(i)}(x_j,x_k)\nonumber \\
&& \times \prod_{l \in \alpha 
\setminus (i,k)} [ P^{(i)}(x_l)] \psi_\alpha(x_\alpha)
\end{eqnarray} 
These functions may be interpreted as generalized 
messages, where the $\mu_{\alpha \to i}(x_i)$ are the familiar ones appearing
in belief propagation.
Putting things together, we may write up to first order:
\begin{eqnarray}
P^{(j)}(x_i) & \approx& \frac{G^{(\alpha)}(x_i)}{\sum_{x_i}G^{(\alpha)}(x_i)} 
\nonumber \\
&& \hspace*{-20mm}-
\frac{\sum_{\beta \in \partial_j \setminus \alpha}\sum_{x_j}
 \nu_{\beta \to j}(x_j,x_i)\prod_{\gamma \in \partial_j
\setminus (\alpha, \beta)}\mu_{\gamma \to j}(x_j)}{
\sum_{x_j}\prod_{\beta \in \partial_j \setminus \alpha}
\mu_{\beta \to j}(x_j)} \nonumber \\
\label{messpass}
\end{eqnarray}
where
\begin{eqnarray}
G^{(\alpha)}(x_i)& = & \left[\prod_{\beta \in \partial_i
\setminus \alpha}\mu_{\beta \to i}(x_i)  \right. \nonumber \\
&& \hspace*{-20mm} \left.  + \sum_{\beta \in \partial_i
\setminus \alpha} \lambda_{\beta \to i}(x_i)\prod_{\gamma \in \partial_i
\setminus (\alpha, \beta)}\mu_{\gamma \to i}(x_i) \right. 
\nonumber \\ 
&& \left. \hspace*{-20mm} +
\sum_{\beta < \gamma \in \partial_i \setminus \alpha}
\rho_{\beta, \gamma \to i}(x_i)\prod_{\eta \in \partial_i \setminus
(\alpha, \beta, \gamma)} \mu_{\eta \to i}(x_i)\right] \nonumber \\
\label{Gcav}
\end{eqnarray}
the true marginals
\begin{eqnarray}
P(x_i) & = & c\sum_{x_{\partial_i}}P^{(i)}(x_{\partial_i})\prod_{\beta \in
\partial_i}\psi_\beta(x_\beta) \nonumber \\
\end{eqnarray}
up to first order read 
\begin{eqnarray}
P(x_i)& \approx& \frac{G(x_i)}{\sum_{x_i}G(x_i)} \\
G(x_i) & = &\left[ \prod_{\beta \in \partial_i}
\mu_{\beta \to i}(x_i) \right. \nonumber \\
&& \left. + \sum_{\beta \in \partial_i}
\lambda_{\beta \to i}(x_i) \prod_{\gamma \in \partial_i \setminus \beta}
\mu_{\gamma \to i}(x_i)  \right. \nonumber \\
&& \hspace*{-10mm}
\left. + \sum_{\alpha < \beta \in \partial_i}\rho_{\alpha,\beta \to i}
(x_i)\prod_{\gamma \in \partial_i \setminus \alpha,\beta} \mu_{\gamma \to i}
(x_i)\right]
\label{Gtot}
\end{eqnarray}
From these equations the pair-interaction case may straightforwardly
be recovered: the $\lambda$ messages do not appear, 
and the remaining messages are given by
\begin{eqnarray}
\mu_{\alpha \to i}(x_i) & = & \sum_{x_k}P^{(i)}(x_k)
\psi_\alpha(x_i,x_k) \\
\rho_{\alpha, \beta \to i}(x_i) & = & \sum_{x_k,x_l}
c^{(i)}(x_k,x_l) \nonumber \\
&&\times \psi_\alpha(x_i, x_k)
\psi_\beta(x_i,x_l) \\
\nu_{\alpha \to i}(x_i, x_j) & = & \sum_{x_k}c^{(i)}(x_j,
x_k)\psi_\alpha(x_i,x_k)
\end{eqnarray}
Note that the messages $\lambda_{\alpha \to i}$, $\rho_{(\alpha,\beta)\to i}$ 
and $\nu_{\alpha \to i}$ should all be ``small'' compared to 
$\mu_{\alpha \to i}$ in absolute sense. If they are not, the method
is bound to fail in this order of approximation. Indeed these messages are not 
normalized, contrarily to the $P^{(i)}(x_j)$. When the ``small'' messages
are actually neglected in total, it is easily seen that one recovers the
belief propagation equations.
\subsection{Complexity issues}
In the above form, we can nicely distinguish the dependence of the
complexity of the algorithm on factor size versus size of the cavities.
The computation of the messages looks exponential in the quadratic factor
size, but this cost may be reduced by storing quantities of the form
\begin{eqnarray}
\phi_\alpha(x_i,x_j) & = & \sum_{x_{\alpha \setminus (i,j)}} \prod_{
k \in \alpha \setminus (i,j)} [P^{(i)}(x_k)] \psi_\alpha(x_\alpha)
\end{eqnarray}
Using these quantities, the computation time scales slightly worse than 
exponential in the factor size.
The dependence on the number of factors in a cavity is found from equation
(\ref{messpass}), from which it is obvious that this dependence is quadratic.

\section{Results: confirming scaling of the error with $N$}

In \cite{MR} the perturbative nature of the expansion on random graphs 
has been confirmed by computing the average of the energy
density in the paramagnetic phase of a spin-glass defined on a random graph.
In particular it has been checked that the first order approximation yields the
$O(1/N)$ correction to the energy that was computed independently through the
replica method.

Currently we report tests of the approximation and the corresponding algorithms on specific instances of random graphs.
We have applied the algorithms CA[0] ({\it i.e.} BP), CA[1] and CA[2] to systems of  binary variables (spin $\sigma_i=\pm 1$, $i \in \{1,\ldots,N\}$) 
described by 
the microstate probability distribution
\begin{eqnarray}
P(\sigma) = \frac{e^{\beta \sum_{i<j}J_{ij}\sigma_i \sigma_j +\sum_i H_i 
\sigma_i}}{
\sum_{\sigma} e^{\beta \sum_{i<j}J_{ij}\sigma_i \sigma_j+\sum_i H_i 
\sigma_i}}
\end{eqnarray}
The nonzero entries of the matrix $J_{ij}$ form a random graph of 
fixed connectivity
equal to three, and we subsequently investigated ferromagnetic interactions
($J_{ij}=1$ for all nonzero entries) and spin-glass interactions 
($J_{ij}=\pm1$ with equal probability for all nonzero entries).

The use of binary variables allows to write the equations (\ref{exactce}) in terms of magnetizations and connected correlation functions (see \cite{MR}), as explained in section \ref{perturbative}, it is assumed that all connected cavity correlations of more than $k+1$ spins vanish when applying the CA[$k-1$] algorithm in the intermediate step of the CA[$k$] algorithm.
\begin{figure}[htb]
\epsfig{file=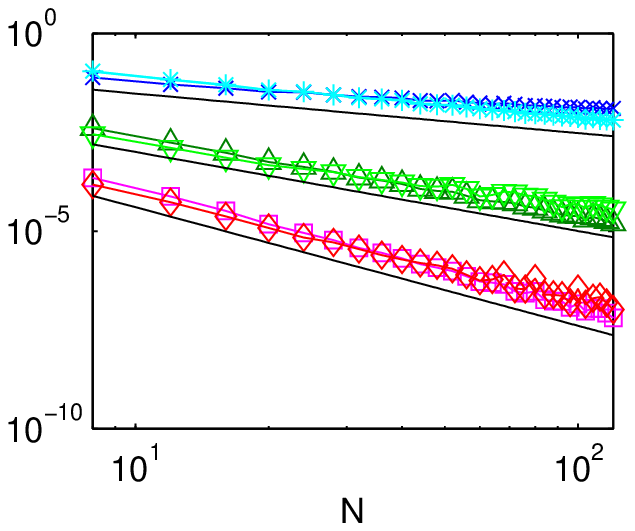,width=8cm}
\epsfig{file=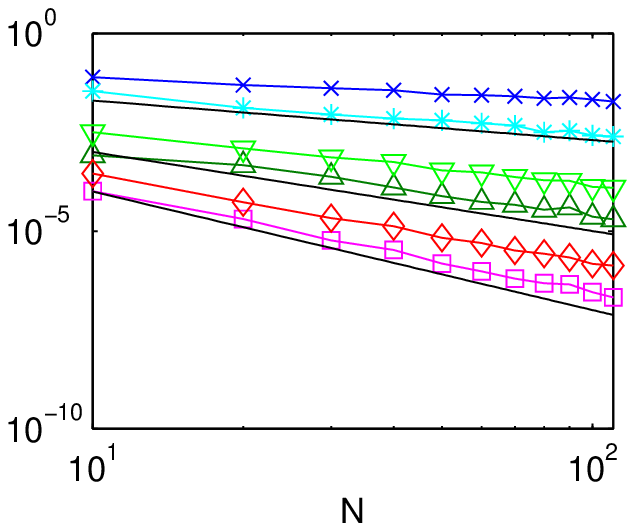,width=8cm}
\caption{Top: ferromagnet on random graph with $\beta=0.3$, fixed connectivity $3$ and $100$ samples. Datapoints represent averages of the errors of the total energy and mean-squared error per link of the estimates of CA[0] (blue, chyan), CA[1] (light green, dark green) and CA[2] (red, magenta), as a function of the sample size $N$, see text. Bottom: same as above, for a spin-glass model, but average is taken in the log-domain (see text).
}
\label{ferro1}
\end{figure}
\begin{figure}
\epsfig{file=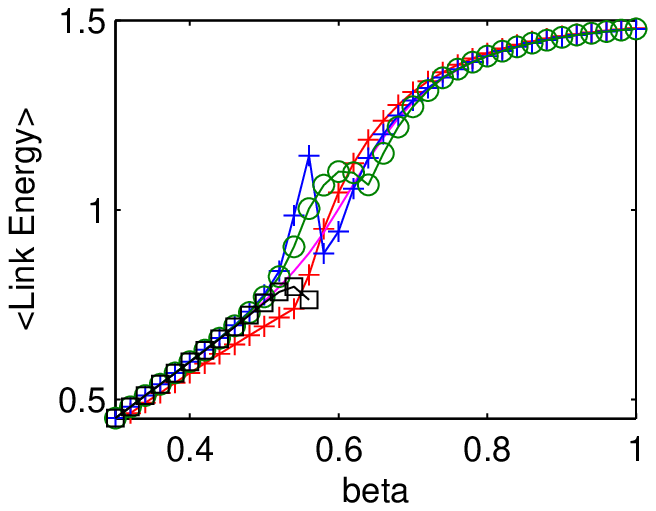,width=8cm}
\epsfig{file=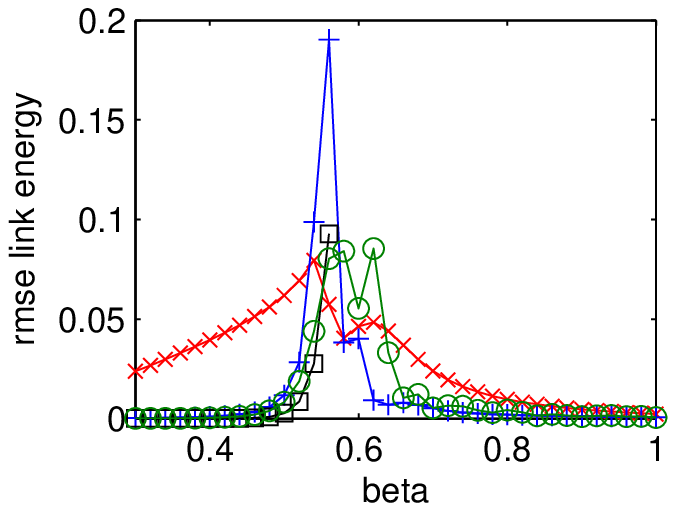,width=8cm}
\caption{
Top: Average energy per link as a function of $\beta$ for an $N=120$ random ferromagnet as obtained exactly via junction tree (magenta), BP or CA[0] (red $\times$), CA[1] with response propagation initialisation (blue $+$), CA[1] initialised with the clamping procedure discussed in this paper (green $\circ$) and CA[2] (only in the paramagnetic regime, black $\square$).
Bottom: Mean square error of link energy, same colour code as top.
Note that in the regime approaching the phase transition the clamping strategy
(green $\circ$) seems to perform better than the response propagation 
procedure.}
\label{ferro2}
%\caption{Same as fig. \ref{spinglass} for a spin-glass model, see text}
%\label{spinglass}
%\caption{Energy}
%\label{energy}
%\caption{Root mean square deviation of energies}
%\label{rootmeansq}
\end{figure}
\begin{figure}[htb]
\epsfig{file=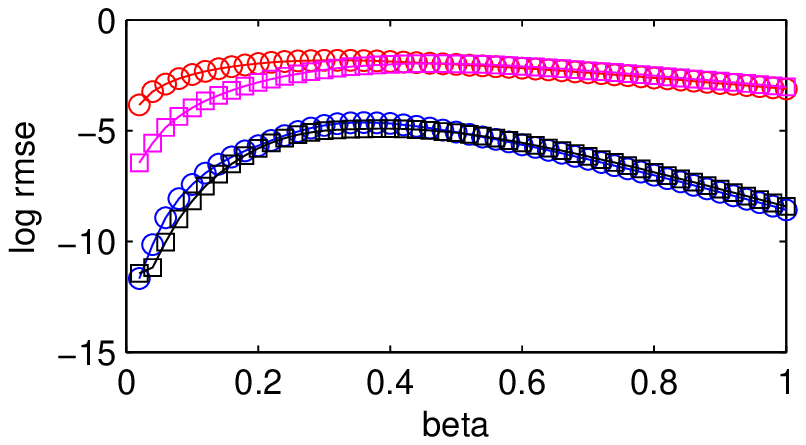,width=7cm,height=5cm}
\epsfig{file=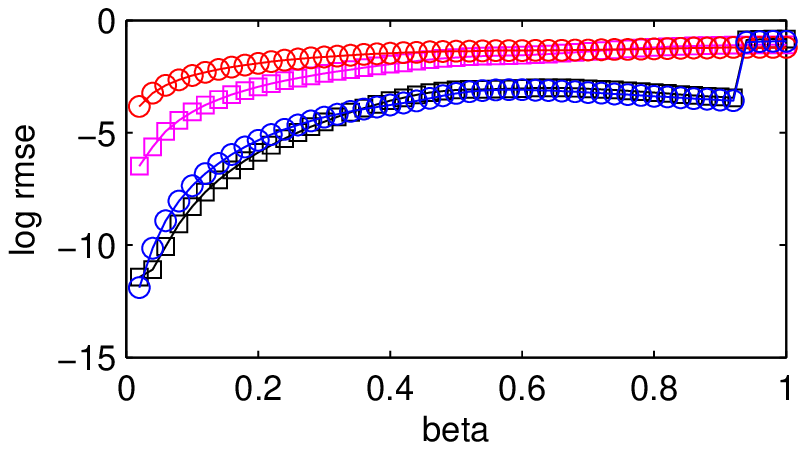,width=7cm,height=5cm}
\caption{Model of ferromagnet on a $k=3$ $N=60$ random graph with 
nonzero magnetizations, random normally distributed
external fields of variance $1$ and average $1$ (top figure) or $0.2$ 
(bottom figure).
Circles correspond to log root mean square error of link energies for BP
(higher, red) versus first order corrected BP (lower, blue). Squares denote 
log root mean square errors of single variable averages for BP (higher, 
magenta) 
and first order corrected BP (lower, black). For average external field
$0.2$ (bottom figure) the first order corrected algorithm does not converge for
$\beta>0.9$. }
\label{magnetized}
\end{figure}
%\vspace*{-0.5cm}
In the top figure (\ref{ferro1}) we report the results for a ferromagnet with $H_i= 0 \ \forall i$ at $\beta=0.3$, corresponding to a paramagnetic phase
(note that for $N\to \infty$ the critical temperature is given by
$\beta_c= \frac{1}{2}\log(3)$).
We compared the various estimates obtained with CA[0], CA[1] and CA[2] with the exact result obtained through a junction tree algorithm \cite{pearl88}, thus we were forced to consider systems sizes up to $N=120$, although the algorithms we are considering can be applied to much larger systems.
For different sizes of the system we plot the average over $100$ random instances of the error of the estimate of the total energy and of the average mean-squared error of the energy of each link. As expected we see that the three algorithms give results of increasing precision, furthermore we see that the error of the BP (CA[0]) total energy scales with the systems size as $1/N$ while those of CA[1] and CA[2] scale respectively as $1/N^2$ and $1/N^3$. 

%\vspace*{-0.5cm}
In the bottom figure (\ref{ferro1}) we report analogous results for a spin-glass again with $H_i=0$ at $\beta=0.3$ and with random interactions $J_{ij}=\pm 1$.
We note that both cases correspond to a paramagnetic region (the critical
temperature for the $N\to \infty$ spin-glass is given by
$\beta_c = \frac{1}{2}\log\left(\frac{\sqrt{2}+1}{\sqrt{2}-1}\right)$), although the algorithm can be applied also in the ferromagnetic region. In the 
paramagnetic region, however, one may exploit the fact that odd moments of 
all (marginal) distributions are zero, significantly reducing the complexity 
of the algorithm. 
The results for the spin glass model naturally display more fluctuations than 
the ones for the ferromagnet, since the interaction values are drawn from a
distribution, whereas for the ferromagnet they are all equal and thus identical
for each of the 100 instances. Since large deviations dominate the average
errors for small error values, we plotted error averages in the log domain 
for the spin glass, i.e. $\exp \langle \log (\Delta E) \rangle$. Although
the correspondence is less convincing than for the ferromagnet, the scaling of
errors roughly follows the same exponents. The deviation from this behaviour
should disappear for larger $N$.
Details of the corresponding equations in the algorithm, using linear response,
are given in appendix 
\ref{App}.% New text:

The figures (\ref{ferro2}) illustrate the $\beta$-dependence, where in the top
figure the total energy is plotted, and in the bottom the root mean square
error of link energies, both as a function of $\beta$, for a ferromagnet with $N=120$. Clearly the 
CA[1] and CA[2] methods outperform the BA in all but a small region around the
``phase transition'', where correlation lengths diverge and
consequently the connected correlation terms blow up. Naturally
perturbative approaches do not result in improved estimates of marginals in
this regime, in
fact the CA[1] and CA[2] methods cease to converge around $\beta=0.55$. Note
that the most difficult part is not estimating the connected correlations, 
which is based on the BA (and BP converges relatively close to the critical
$\beta$), 
but the adapted update equations for $P^{(i)}(x_j)$,
equation (\ref{exactce}) in CA[1] (equation (\ref{directcavm}) in the 
appendix), do not converge. In general one might 
be able to optimize an update scheme for these equations, which we have 
not attempted here.

Apart from problems around the critical value of $\beta$, for larger $\beta$
the inversion symmetry in the model is broken by the approximation algorithms,
that consequently disregard the mirror-state free energy valley. In these
small-scale models, this does affect the results as can be seen from 
figure \ref{ferro2}.

When the symmetry of the model is already broken by a sufficiently large
external field, a situation which is common in statistical inference 
applications,
where the external fields may originate from Bayesian
priors or represent evidence from measurement data (see e.g. \cite{mackay}), 
this phenomenon does not occur. 
The CA[1] algorithm 
consistently improves the marginal estimates over the whole range of $\beta$,
as illustrated in figure \ref{magnetized}. When the average external field
is relatively small, symmetry breaking might again prevent convergence
(figure \ref{magnetized} bottom for $\beta>0.9$).

\begin{figure}[htb]
\epsfig{file=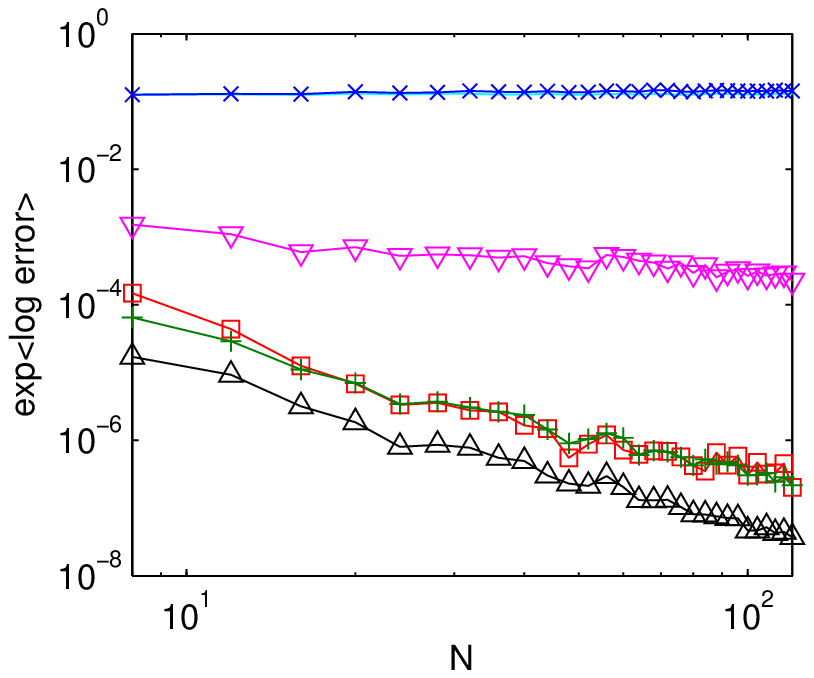,width=8cm,height=5cm}
\epsfig{file=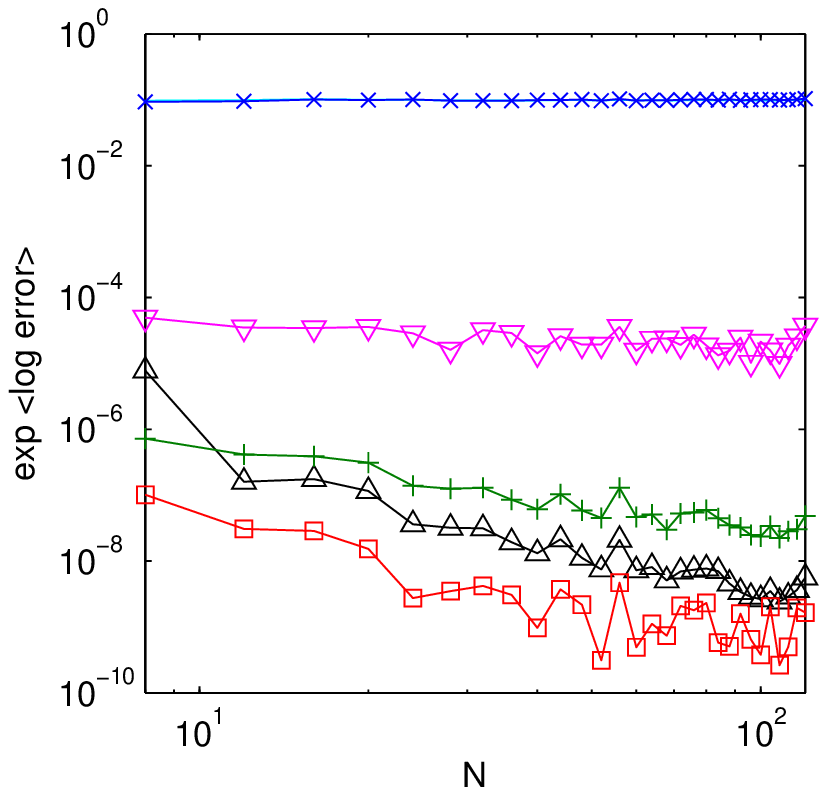,width=7cm, height=5cm}
\caption{Scaling behaviour for a ferromagnet
in a broken symmetry magnetized regime caused by normally 
distributed external fields $H_i$ of variance $1$ and average $0.5$ (top)
versus $1$ (bottom). Blue $(\times)$: root mean square error in BP
correlations, magenta $(\bigtriangledown)$, root mean square error in BP site
magnetizations. Red $(\square)$: root mean square error in CA[1] 
correlation, green $(+)$, root mean square error in CA[1] site 
magnetizations. Black $(\bigtriangleup)$: error in CA[1] mean energy per link.
All data are averages over 100 instances of a random graph of connectivity 3.}
\label{magscaling}
\end{figure}

In the ``magnetized'' regime of this model, one may do a similar scaling 
analysis as displayed
in figure \ref{ferro1}. Results are reported in figure \ref{magscaling}
for a model with
ferromagnetic interactions and different values of the external field
average, where
we plot the error of the first order CA[1] algorithm as a function of $N$.
Although the results display more fluctuations, the behaviour is similar, 
in the sense that one observes on average a scaling with 
showing that indeed as long as the parameters correspond to regions not 
in the vicinity of a phase transition, and the correlation
lengths remain typically small compared to the loop length,
the approach is promising. 

\section{Discussion}
The implemented approach is intrinsically perturbative around the BA, in the sense that the BA gives accurate results if the correction terms in eq. (\ref{exactce}) are small and therefore it is natural to guess that CA[1] will produce better results. At the same time if the corrections turn out not to be small, this hints at poor BP estimates, and the whole approach is in trouble (see bottom figure \ref{ferro2}). Furthermore we cannot compute CA[1] if BP does not converge. However we recall that {\it any} algorithm can be used  as the starting point CA[0] of the sequence of approximations. 
In conclusion we expect that {\it whenever BP converges and yields good estimates, CA[$k$] yields a series of approximations of increasing precision. In particular for graphical models defined on random graphs where small loops are rare, CA[$k$] gives estimates with an error $O(1/N^{k+1})$ and with computational complexity $N^{k+1}$.}
Note that this last class of models includes some of the most important present-date error-correcting codes for which the decoding scheme is BP. The reason why  BP is so efficient in these cases is precisely that in the corresponding graphical models small loops are rare. Therefore we expect that by the application of CA[$k$] the marginals can be computed with any precision in polynomial time. It is important to realize that this does not completely solve the problem of the $1/N$ finite-size effect in error-correcting-codes, indeed even if we know the exact marginals, there is still the possibility that some of them are not consistent with the encoded original message. 

\section{Relation with other approaches}

The previous comments should help the reader to understand what is the natural context of the present approach and to clarify the relationship with different approaches.
A well known generalisation of the BA is Kikuchi's cluster variation method (CVM) \cite{Kik} which is particularly suitable for finite-dimensional models and in general for models where many small loops are present, indeed this approach amounts to treating loops up to a certain length exactly.
On the other hand on random graphs the corrections to the BA are determined not by small loops (which are rare) but by many large loops. The CVM does not apply to such cases since, in order to include the effect of the large loops, the size of the basic clusters that it treats exactly should be of the order of the total system size, with prohibitive computational complexity.
On the other hand it is natural to expect that CVM performs much better than CA[$k$] on graphical models defined on structures with many small loops like lattices. 
Thus the cavity approaches are complementary to CVM, in the sense that both methods have their own well-defined range of applications, although one can imagine applications that could be best studied through a mixture of them.

In a recent publication Chertkov and Chernyak (CC) \cite{CC} obtained the free energy of a generic graphical model as an expansion around the BA written in terms of diagrams corresponding to sub-graphs with one loop, two loops and so on. In spite of their claim that this represents an improvement with respect to the approach presented here we believe that the two approaches have different motivations and capabilities.
The present approach addresses the problem of improving the computation of marginals with polynomial agorithms for models defined on random graphs (with error correcting codes being a notable example of this class of models) and it is as yet not clear if similar results are achievable within the CC approach. Indeed we know that the $1/N$ corrections computed by CA[1] with $N^2$ complexity are determined by exponentially many large loops (each one yielding a small contribution), therefore  it seems likely that in order to obtain results of the quality of CA[1] ({\it i.e.} the $1/N$ corrections) one should consider all graphs with one loop in the CC expansion, yielding an exponential number of terms, which is computationally prohibitive unless some resummation scheme is supplemented. Recently (\cite{Vicenc}) an algorithm was tested based on truncation of the series, which may work in cases where one is able to identify the most important loops that contribute to the BP error, when there are not too many.

In a very recent paper \cite{joris}, a number of different algorithms based on similar ideas as the above have been described, and have been applied to some real-world problems. Given an estimate of the cavity distributions, the update relations in \cite{joris} are based on an adjustment of external fields, (keeping higher order interactions in the cavity distribution fixed, whereas, we keep the higher order connected correlations fixed). Although the connection with higher order improvements in their scheme is lost, the algorithms are sometimes easier to implement in the first order case, if the connectivity of the graph is not too large. 

Some open problems of the present approach include the computation of the $1/N$ corrections to the free energy (currently we know only how to improve the marginals, therefore we have access only to corrections to local quantities such as the magnetization and the energy) and the extension to the spin-glass phase with the inclusion of Replica-Symmetry-Breaking effects.

\section{Acknowledgments}
We thank Joris Mooij for useful discussions and BW acknowledges STW
for funding.

\appendix
\section{CA[1] and CA[2] update equations for connectivity 3.}
\label{App} 
\subsection{CA[1] updates}
The update equations for the first moment $M^{(i)}_j$ of a cavity marginal 
$P^{(i)}(x_j)$
may be written out in terms of connected correlation functions, i.e.
$M^{(i)}_k=\sum_{x_k} c^{(i)}(x_k)x_k$, 
$C^{(i)}_{kl}=\sum_{x_k, x_l} c^{(i)}(x_k,x_l)x_k x_l$ and
$C^{(i)}_{klm}=\sum_{x_k, x_l, x_m} c^{(i)}(x_k,x_l,x_m)x_k x_l x_m$, (see 
\cite{MR}):
\begin{eqnarray}
M_j^{(i)} & = & \frac{t(H_j)K_{ij} + \sum_{l\in \partial j \setminus i}
t_{jl}M_l^{(j)}}
{K_{ij}+ t(H_j)\sum_{l\in \partial j \setminus i}
t_{jl}M_l^{(j)}}\nonumber \\
&&- \frac{L_{ij} 
+ t(H_i)\sum_{k\in \partial i \setminus j} t_{ik}C_{jk}^{(i)}}{
K_{ji} + t(H_i)\sum_{k\in \partial i\setminus j}
t_{ik}M_k^{(i)}}
\label{directcavm}
\end{eqnarray}
where
\begin{eqnarray}
K_{ij}&=&1+\sum_{k<l \in \partial j\setminus i}t_{jk}t_{jl}
[M_k^{(j)}M_l^{(j)}+C_{kl}^{(j)}] \\
L_{ij}&=& \sum_{k<l \in \partial i \setminus j} t_{ik}t_{il}[C_{jl}^{(i)}
M_k^{(i)}+ C_{jk}^{(i)}M_l^{(i)} + C_{jkl}^{(i)}] \qquad
\end{eqnarray}
The solution of these equations leads to the moment of the true marginal 
$P(x_i)$ via
\begin{eqnarray}
M_i & = &\frac{T_{\rm odd}^{(i)}+t(H_i)T_{\rm even}^{(i)}}
{t(H_i)T_{\rm odd}^{(i)} + T_{\rm even}^{(i)}}
\nonumber \\
T_{\rm odd}^{(i)} &=&
\sum_{l\in \partial_i} t_{il}M_l^{(i)}+ 
\sum_{(l,k,m) \in \partial i} t_{il}t_{ik}t_{im}
M_l^{(i)}C_{km}^{(i)} 
 \nonumber \\
&&+\sum_{l<k<m \in \partial_i} t_{il}t_{ik}t_{im}[M_l^{(i)}M_k^{(i)}
M_m^{(i)}+C_{lkm}^{(i)}]\nonumber \\
T_{\rm even}^{(i)}& = &
\left[1+\sum_{k<l \in \partial_i}t_{il}t_{ik}[M_l^{(i)}M_k^{(i)}+
C_{lk}^{(i)}]\right]
\label{truemags}
\end{eqnarray}
Correspondingly, the nearest neighbour correlations read
\begin{eqnarray}
\sum_{x_i, x_j} P(x_i,x_j)x_ix_j =  
\frac{F_{ij}}{T^{(i)}_{\rm even}+t(H_i)T^{(i)}_{\rm odd}} 
\end{eqnarray}
where
\begin{eqnarray}
F_{ij}= t(H_i)[ M_j^{(i)} + L_{ij}]
+ \sum_{l \in \partial_i \setminus j} t_{il} [C_{jl}^{(i)}
+M_j^{(i)} M_{l}^{(i)}] \nonumber \\
+ 
t_{ij}\left[K_{ji}
+ t(H_i)\sum_{l\in \partial i \setminus j} t_{il}M_l^{(i)}
\right]
\nonumber \\
+ t(H_i)\sum_{l<k \in \partial i \setminus j}
t_{ik}t_{il}M_j^{(i)}[M_l^{(i)}M_k^{(i)}+C_{lk}^{(i)}] \qquad
\label{truecors}
\end{eqnarray}
In the CA[1] approximation, the two-point connected correlations are estimated
by some algorithm, possibly CA[0] (another option is to use response 
propagation, see \cite{MR}), and the three-point connected correlations
are neglected. 

\subsection{CA[2] updates}
The CA[2] algorithm in turn uses improved 
estimates of the two-point connected correlations of which the accuracy
corresponds to CA[1], together with CA[0] (or response propagation) three-point
estimates. 
We used response propagation to compute the CA[1] accurate 
two-point cavity connected correlations. This implies we exploit
\begin{equation}
C^{(j)}_{ik} = \beta^{-1}\frac{\partial M^{(j)}_i}{\partial H_k}
\end{equation} 
but $M_i^{(j)}$ is computed with CA[1] accuracy, i.e., from equation
(\ref{truemags}) on the graph from which variable $j$ has been removed.
This may be achieved by simply taking the derivative of the right hand
side of equation (\ref{truemags}). In this expression, we encounter
$\frac{\partial M_k^{(i)}}{\partial H_l}$ and 
$\frac{\partial C_{kl}^{(i)}}{\partial H_m}$. The first may be found from the
iterative equation resulting from taking the derivative of (\ref{directcavm}),
the second may be estimated with a CA[0] or response propagation algorithm, 
since it is of the order of $C_{klm}^{(i)}$.

In the paramagnetic phase of pair interaction networks without external field, 
simplifications occur since we may exploit the fact
that odd moments of distributions are zero. Consequently, all terms $M_i$,
$M_i^{(j)}$, $C_{ikl}^{(j)}$ and $\frac{\partial C_{jk}^{(i)}}{\partial H_l}$ 
vanish and the recursive update relations for the derivatives of 
(\ref{directcavm}) reduce to
\begin{eqnarray}
\frac{\partial M_l^{(j)}}{\partial H_n}
= \left[ \frac{\sum_{k\in \partial l\setminus j} t_{lk}\frac{\partial 
M_k^{(l)}}{\partial H_n}}{1+\sum_{k<r \in \partial l \setminus j }
t_{lk}t_{lr}C_{kr}^{(l)}} + \delta_{ln}\beta\right] \nonumber \\
- \frac{\sum_{k<r \in \partial j \setminus l}t_{jk}t_{jr}
\left[ \frac{\partial M_r^{(j)}}{\partial H_n} C_{lk}^{(j)}+ 
\frac{\partial M_k^{(j)}}{\partial H_n} C_{lr}^{(j)}\right]
}
{1+ \sum_{k<r \in \partial j\setminus l} t_{jk}t_{jr}C_{kr}^{(j)}}
\nonumber \\
+ \delta_{jn}\frac{\beta\sum_{k\in \partial j \setminus l}t_{jk}C_{lk}^{(j)}}
{1+ \sum_{k<r \in \partial j\setminus l} t_{jk}t_{jr}C_{kr}^{(j)}}
\end{eqnarray}
The solution of these equations is to be substituted in 
\begin{equation}
\frac{\partial M_j}{\partial H_n} = 
\frac{ \sum_{l \in \partial j} t_{lj} \frac{\partial M_l^{(j)}}{\partial H_n}
}{ 1+\sum_{k<l \in \partial j} t_{jk}t_{jl} C_{lk}^{(j)}}
\end{equation}
on the graph without variable $i$, yielding a CA[1] computation of 
$C^{(i)}_{jn}$. This results in improved CA[2] estimates of correlations via 
equation (\ref{truecors}) which simplifies greatly due to the vanishing of 
odd moments, i.e.
\begin{eqnarray}
\sum_{x_i,x_j}P(x_i,x_j)x_i x_j =  t_{ij} + \frac{\sum_{k \in \partial_i 
\setminus j}t_{ik}C_{jk}^{(i)}}
{1+\sum_{k<l \in \partial_i}t_{ik}t_{il}C_{kl}^{(i)}} \ \
\end{eqnarray}

\end{document}